\documentclass[aps,preprint]{revtex4}%
\usepackage{amsfonts}
\usepackage{amsmath}
\usepackage{amssymb}
\usepackage{graphicx}%
\begin{document}
\preprint{CTP-SCU/2014005}
\preprint{CAS-KITPC/ITP-442}
\title{Deformed Hamilton-Jacobi Method in Covariant Quantum Gravity Effective Models}
\author{Mu Benrong$^{a,b}$}
\email{mubenrong@uestc.edu.cn}
\author{Peng Wang$^{b}$}
\email{pengw@scu.edu.cn}
\author{Haitang Yang$^{b,c}$}
\email{hyanga@scu.edu.cn}
\affiliation{$^{a}$School of Physical Electronics, University of Electronic Science and
Technology of China ,Chengdu, 610054, China}
\affiliation{$^{b}$Center for Theoretical Physics, College of Physical Science and
Technology, Sichuan University, Chengdu, 610064, PR China}
\affiliation{$^{c}$Kavli Institute for Theoretical Physics China (KITPC), Chinese Academy
of Sciences, Beijing 100080, P.R. China}

\begin{abstract}
We first briefly revisit the original Hamilton-Jacobi method and show that the
Hamilton-Jacobi equation for the action $I$ of tunnelings of a fermionic
particle from a charged black hole can be written in the same form as that of
a scalar particle. For the low energy quantum gravity effective models which
respect covariance of the curved spacetime, we derive the deformed
model-independent KG/Dirac and Hamilton-Jacobi equations using the methods of
effective field theory. We then find that, to all orders of the effective
theories, the deformed Hamilton-Jacobi equations can be obtained from the
original ones by simply replacing the mass of emitted particles $m$ with a
parameter $m_{eff}$ that includes all the quantum gravity corrections.
Therefore, in this scenario, there will be no corrections to the Hawking
temperature of a black hole from the quantum gravity effects if its original
Hawking temperature is independent of the mass of emitted particles. As a
consequence, our results show that breaking covariance in quantum gravity
effective models is a key for a black hole to have the remnant left in the evaporation.

\end{abstract}
\keywords{}\maketitle
\tableofcontents



\section{Introduction}

Hawking radiation is a theoretical argument proposed by Stephen Hawking for
the existence of thermal radiation emitted by black holes. The standard
Hawking formula was first obtained in the frame of quantum field theory in
curved spacetime\cite{Hawking1975}. Afterward, there are various methods for
deriving Hawking radiation and calculating Hawking temperature. Among them is
a semiclassical method of modeling Hawking radiation as a tunneling effect.
This method was first proposed by Kraus and
Wilczek\cite{Wilczek1995-1,Wilczek1995-2}. They employed the dynamical
geometry approach to calculate the imaginary part of the action for the
tunneling process of s-wave emission across the horizon and related it to the
Boltzmann factor for the emission at the Hawking temperature. Due to back
reaction effects included, this procedure gives a correction to the standard
Hawking temperature formula, which speeds up the process of black holes'
evaporation. An alternative way to calculate the imaginary part of the action
is the Hamilton-Jacobi method\cite{Angheben2005,Kerner2006}. Neglecting the
self-gravitation, this method assumes the action of an emitted particle
satisfies the relativistic Hamilton-Jacobi equation. Taking the symmetries of
the metric into account, one can adopt an appropriate ansatz for the form of
the action. Solving the Hamilton-Jacobi equation turns out to recover the
standard Hawking temperatures.

However, the original Hamilton-Jacobi method proposed in
\cite{Angheben2005,Kerner2006} is only confined to the semiclassical
approximation and carried out in the framework of the classical general
relativity. Beyond the original Hamilton-Jacobi method, there could be some
corrections. For simplicity, we use the case of emission of massless scalar
with energy $E$ from the Schwarzschild black hole with mass $M$ to illustrate
these corrections. In this paper, we set $G=c=1$, where the Planck constant
$\hbar$ is of the order of square of the Planck Mass $M_{p}$. In these units,
the black hole's horizon $r_{H}=2M$. The potential corrections to the original
Hamilton-Jacobi method are given as follows:

\begin{enumerate}
\item[$\left(  a\right)  $] Back-reaction effects, which gives the correction
$\sim\frac{E}{M}$. In the Hamilton-Jacobi method, we assume that the metric is
fixed and the field is free. Thus, back-reaction effects are disregarded.
However, in Parikh and Wilczek method\cite{Wilczek1995-1,Wilczek1995-2}, there
are back-reaction effects to ensure energy conservation during the emission of
a particle via tunneling through the horizon. These corrections lead to
non-thermal corrections to the black-hole radiation spectrum.

\item[$\left(  b\right)  $] Higher order WKB corrections. Using WKB
approximation method, the lowest order of the equation of motion describing a
particle moving in the black hole gives the Hamilton-Jacobi equation. The WKB
approximation breaks down when the de Broglie wavelength of the particle,
$\lambda_{p}\sim\frac{\hbar}{E}$, becomes comparable to the horizon of black
hole, $r_{H}$. Therefore, the ratio $\frac{\lambda_{p}}{r_{H}}\sim\frac{\hbar
}{EM}$ controls the amount of higher order WKB corrections. However, several
authors\cite{Yale2011,Mitra2009,Wang2010} argued that the tunneling method
yields no higher-order corrections to the Hawking temperature.

\item[$\left(  c\right)  $] Quantum gravity corrections. The original Hawking
formula predicts the complete evaporation of black holes, which leads to the
black hole information paradox (for reviews see \cite{Preskill1992,Russo2005}%
). Solving the paradox needs the breakdown of the semiclassical description at
the Planck scale. A theory of quantum gravity must be used to describe the
final state of black hole evaporation. Although a number of quantum gravity
theories are proposed, there is still no complete and consistent quantum
theory of gravity. Therefore, in the absence of a full quantum description of
the black hole evaporation, one uses effective models to describe the quantum
gravitational behavior. For example, various theories of quantum gravity, such
as string theory, loop quantum gravity and quantum geometry, predict the
existence of a minimal length. An effective model to realize this minimal
length is the generalized uncertainty principle (GUP). Adler \textit{et
al.}\cite{Adler2001}\textit{ }showed that incorporating the GUP into the
derivation of Hawking temperature could predict the existence of black hole
remnants. Recently, the GUP modified Hamilton-Jacobi equation for fermions in
curved spacetime have been introduced and the corrected Hawking temperatures
have been
derived\cite{Chen:2013pra,Chen2014-1,Chen2014-2,Chen:2013kha,Ren2014,Chen:2014xsa}%
. While in the Parikh-Wilczek tunneling picture, the Planck-scale corrections
to the black-hole radiation spectrum, due to the GUP\cite{Nozari2012}, the
noncommutative geometry\cite{Banerjee2008,Zhang2011}, and, a modification of
the energy-momentum dispersion relation caused by the quantum gravity
effects\cite{Arzano2005}, have also been studied.\ 
\end{enumerate}

The existence of the minimal length implied by the quantum gravity theory
could lead to Planck scale departures from Lorentz symmetry. The
deformed\ commutation relations and modified dispersion relations proposed as
low-energy quantum gravity effects could deform or break the Lorentz symmetry.
Effective models incorporating with such deformed\ commutation relations or
dispersion relations hence are not covariant in
spacetime\cite{Hossenfelder2003,Camacho2003,Maziashvili2013,Berger2011}. The
deformed non-covariant Hamilton-Jacobi equation have been discussed
in\cite{Chen:2013pra,Chen2014-1,Chen2014-2,Chen:2013kha,Ren2014,Chen:2014xsa}.
Although covariance is not required for constructing low-energy quantum
gravity effective models, there are indeed some attempts to introduce
covariant models\cite{Tkachuk2006,Kober2010}. The gauge theories with
incorporation of the GUP in \cite{Kober2010} are both covariant and gauge
invariant. In addition, the Klein-Gordon equation described by Quesne-Tkachuk
Lorentz-covariant deformed algebra has been given in \cite{Moayedi2010}. So
far, the quantum gravity corrections to the Hamilton-Jacobi method in these
covariant effective models have not been discussed. Thus, we will investigate
the quantum gravity corrections to the Hamilton-Jacobi method in the covariant
low energy quantum gravity effective models in a model-independent way in this
paper. To achieve this aim, we will work with an effective theory that is
valid below a scale $\Lambda$, at which the quantum gravity effects becomes
important. As usual, we assume $\Lambda\sim M_{p}$. We also assume that the
effective theory respects covariance of the curved spacetime and $U\left(
1\right)  $ gauge invariance of the charged black hole. Since the particles
tunneling through the horizon are treated as free particles in the original
Hamilton-Jacobi method, a natural assumption is that there are no
self-interacting effective operators in the effective field theory.

Using the effective theory method, we will show that there are no corrections
to the Hawking temperature of a black hole from the quantum gravity effects in
the covariant effective models if its original Hawking temperature is
independent of the mass of emitted particles. For these purposes, this paper
is organized as follows. In Section \ref{Sec:HJ}, we briefly review the
original Hamilton-Jacobi method and show the fermionic Hamilton-Jacobi
equation for the action $I$ can be written in the same form of the scalar one.
After constructing scalar(fermionic) effective field theories respecting
covariance of coordinate spacetime, we derive the deformed model-independent
KG/Dirac and Hamilton-Jacobi equations in Section \ref{Sec:HJG}. Section
\ref{Sec:Con} is devoted to our discussion and conclusion.

\section{Hamilton-Jacobi Method}

\label{Sec:HJ}

In this section, we briefly review how to calculating the imaginary part of
the action making use of the Hamilton-Jacobi equation \cite{Angheben2005}.
This semiclassical method models Hawking radiation as a tunneling through the
horizon. Using the WKB approximation, the tunneling probability for the
classically forbidden trajectory through the horizon is given by:%
\begin{equation}
\Gamma\propto\exp\left(  \frac{-2\operatorname{Im}I}{\hbar}\right)  ,
\end{equation}
where $I$ is the classical action of the trajectory. One can relate $\Gamma$
to the Boltzmann factor for the emission from the black hole to get Hawking temperature.

\subsection{Scalar Field}

The equation satisfied by the scalar field is%
\begin{equation}
D^{\mu}D_{\mu}\phi+\frac{m^{2}}{\hbar^{2}}\phi=\left(  \nabla^{\mu}+\frac
{ie}{\hbar}A^{\mu}\right)  \left(  \nabla_{\mu}+\frac{ie}{\hbar}A_{\mu
}\right)  \phi+\frac{m^{2}}{\hbar^{2}}\phi=0\text{,} \label{eq:KG}%
\end{equation}
where $\nabla_{\mu}$ is the covariant derivative of the black hole and
$A_{\mu\text{ }}$is its electromagnetic potential. Making the ansatz for
$\phi$ which is%
\begin{equation}
\phi=\exp\left(  \frac{iI}{\hbar}\right)  ,
\end{equation}
and substituting it into eqn. $\left(  \ref{eq:KG}\right)  $, one expands eqn.
$\left(  \ref{eq:KG}\right)  $ in powers of $\hbar$ and finds to the lowest
order%
\begin{equation}
\left(  \partial^{\mu}I+eA^{\mu}\right)  \left(  \partial_{\mu}I+eA_{\mu
}\right)  -m^{2}=0. \label{eq:Hamilton-JacobiS}%
\end{equation}
Eqn. $\left(  \ref{eq:Hamilton-JacobiS}\right)  $ is just the Hamilton-Jacobi
equation satisfied by a scalar particle of mass $m$ moving in the black hole
with the electromagnetic potential $A_{\mu}$. The solution to eqn. $\left(
\ref{eq:Hamilton-JacobiS}\right)  $ is the action of the scalar's classically
forbidden trajectory through the horizon.

To illustrate how the Hamilton-Jacobi method works, we consider Hawking
radiation in the $\left(  1+1\right)  $ dimensional Schwarzschild black hole
with line element (with $c=G=1$)%
\begin{equation}
ds^{2}=\left(  1-\frac{2M}{r}\right)  dt^{2}-\left(  1-\frac{2M}{r}\right)
^{-1}dr^{2}.
\end{equation}
Thus, eqn. $\left(  \ref{eq:Hamilton-JacobiS}\right)  $ becomes%
\begin{equation}
\left(  1-\frac{2M}{r}\right)  ^{-1}\left(  \partial_{t}I\right)  ^{2}-\left(
1-\frac{2M}{r}\right)  \left(  \partial_{r}I\right)  ^{2}-m^{2}=0,
\end{equation}
where we use $A_{\mu}=0$ for the Schwarzschild black hole. Using the method of
separation of variables, we find that the solution to the above equation is%
\begin{equation}
I_{\pm}=-Et\mp\int\frac{dr}{1-\frac{2M}{r}}\sqrt{E^{2}-m^{2}\left(
1-\frac{2M}{r}\right)  },
\end{equation}
where $E$ is a constant and $+\left(  -\right)  $ corresponds to the outgoing
(ingoing) solutions. Choosing the contour to lie in the upper complex plane,
one gets that the imaginary part of $I$ is
\begin{equation}
\operatorname{Im}I_{\pm}=\mp\int_{2M-\epsilon}^{2M+\epsilon}\frac{dr}%
{1-\frac{2M}{r}}\sqrt{E^{2}-m^{2}\left(  1-\frac{2M}{r}\right)  }=\pm2\pi EM.
\end{equation}
Thus, the tunneling rate is%
\begin{equation}
\Gamma=\frac{P_{(emission)}}{P_{(absorption)}}=\frac{\exp\left(
-2\operatorname{Im}I_{+}\right)  }{\exp\left(  -2\operatorname{Im}%
I_{-}\right)  }=exp(-8\pi ME).
\end{equation}
Comparing the above equation with the Boltzmann factor at the Hawking
temperature near the event horizon gives%
\begin{equation}
T=\frac{1}{8\pi M}.
\end{equation}

\subsection{Fermion Field}

In curved spacetime, the Dirac equation for a spin-$1/2$ fermion with an
electromagnetic field $A_{\mu}$ takes on the form as%
\begin{equation}
i\gamma_{\mu}\left(  \partial^{\mu}+\Omega^{\mu}+\frac{ie}{\hbar}A_{\mu
}\right)  \psi-\frac{m}{\hbar}\psi=0, \label{eq:Dirac}%
\end{equation}
where $\Omega_{\mu}\equiv\frac{i}{2}\omega_{\mu}^{\text{ }ab}\Sigma_{ab}$,
$\Sigma_{ab}$ is the Lorentz spinor generator, $\omega_{\mu}^{\text{ }ab}$ is
the spin connection and $\left\{  \gamma_{\mu},\gamma_{\nu}\right\}
=2g_{\mu\nu}$. The Greek indices are raised and lowered by the curved metric
$g_{\mu\nu}$, while the Latin indices are governed by the flat metric
$\eta_{ab}$. To obtain the Hamilton-Jacobi equation for a fermion, the ansatz
for $\psi$ is assumed as
\begin{equation}
\psi=\exp\left(  \frac{iI}{\hbar}\right)  v, \label{eq:fermionansatz}%
\end{equation}
where $v$ is a slowly varying spinor amplitude. Substituting eqn. $\left(
\ref{eq:fermionansatz}\right)  $ into eqn. $\left(  \ref{eq:Dirac}\right)  $,
we find to the lowest order of $\hbar$%
\begin{equation}
\gamma_{\mu}\left(  \partial^{\mu}I+eA^{\mu}\right)  v=-mv,
\label{eq:Hamilton-JacobiF}%
\end{equation}
which is the Hamilton-Jacobi equation satisfied by a fermion particle of mass
$m$ moving in the black hole with the electromagnetic potential $A_{\mu}$.
Solving eqn. $\left(  \ref{eq:Hamilton-JacobiF}\right)  $ gives us the
classical action $I$. Multiplying both sides of eqn. $\left(
\ref{eq:Hamilton-JacobiF}\right)  $ from the left by $\gamma_{\nu}\left(
\partial^{\nu}I+eA^{v}\right)  \,$and then using eqn. $\left(
\ref{eq:Hamilton-JacobiF}\right)  $ to simplify the RHS, one gets%
\begin{equation}
\gamma_{\nu}\left(  \partial^{\nu}I+eA^{v}\right)  \gamma_{\mu}\left(
\partial^{\mu}I+eA^{\mu}\right)  v=m^{2}v.
\end{equation}
Manipulating the LHS of the above equation by using $\left\{  \gamma_{\mu
},\gamma_{\nu}\right\}  =2g_{\mu\nu}$, we have%
\begin{equation}
\left[  \left(  \partial^{\mu}I+eA^{\mu}\right)  \left(  \partial_{\mu
}I+eA_{\mu}\right)  -m^{2}\right]  v=0.
\end{equation}
Since $v$ is nozero, the Hamilton-Jacobi equation satisfied by the classical
action $I$ for a fermion finally becomes%
\begin{equation}
\left(  \partial^{\mu}I+eA^{\mu}\right)  \left(  \partial_{\mu}I+eA_{\mu
}\right)  -m^{2}=0,
\end{equation}
which is the same as the Hamilton-Jacobi equation for a scalar with the same
mass $m$, namely eqn. $\left(  \ref{eq:Hamilton-JacobiS}\right)  $.

\section{ Hamilton-Jacobi Equation With Incorporation of Quantum Gravity
Effects}

\label{Sec:HJG}

To incorporate the quantum effects into the original Hamilton-Jacobi Method,
we need to derive the KG/Dirac equations in low energy quantum gravity
effective models. In this section, we will calculate the deformed KG/Dirac and
Hamilton-Jacobi Equation with incorporating quantum gravity effects in a
covariant way by making using of effective theory Method.

\subsection{ Model-Independent Covariant Deformed KG/Dirac Equations}

To set notation, the effective Lagrangian involving the scalar field $\phi$
(fermion field $\psi$) in the background of a $\left(  D+1\right)
$-dimensional black hole with the electromagnetic potential $A_{\mu}$ is given
by%
\begin{equation}
\mathcal{L}_{eff}^{s\left(  f\right)  }=%
{\displaystyle\sum\limits_{n,j}}
\frac{C_{n,j}^{s\left(  f\right)  }}{\Lambda^{n-\left(  D+1\right)  }%
}\mathcal{O}_{n,j}^{s\left(  f\right)  },
\end{equation}
where $s\left(  f\right)  $ denotes the scalar (fermion), the $n\geq$ $D+1$
denotes the operator dimension, $j$ runs over all independent operators of a
given dimension. The lowest dimensional operator with $n=D+1$ is the original
free field Lagrangian in curved spacetime with the electromagnetic potential
$A_{\mu}$
\begin{align}
\mathcal{O}_{D+1}^{s}  &  =-\phi^{+}\left(  D^{\mu}D_{\mu}+\frac{m^{2}}%
{\hbar^{2}}\right)  \phi,\\
\mathcal{O}_{D+1}^{f}  &  =\bar{\psi}\left[  iD_{\mu}^{f}\gamma^{\mu}-\frac
{m}{\hbar}\right]  \psi,
\end{align}
where $D_{\mu}\equiv\nabla^{\mu}+\frac{ie}{\hbar}A^{\mu}$, $D_{\mu}^{f}%
\equiv\partial_{\mu}+\Omega_{\mu}+\frac{ie}{\hbar}A_{\mu}$, and $m$ is the
mass of the particle. For the fermion field $\psi$, the basis of independent
effective operators with $n=D+2$ is given by%
\begin{gather}
\mathcal{O}_{D+2,1}^{f}=\hbar\bar{\psi}\left(  \gamma^{\nu}D_{\nu}^{f}\right)
\left(  \gamma^{\mu}D_{\mu}^{f}\right)  \psi,\text{ }\mathcal{O}_{D+2,2}%
^{f}=\hbar\bar{\psi}D^{f,\mu}D_{\mu}^{f}\psi,\nonumber\\
\mathcal{O}_{D+2,3}^{f}=im\bar{\psi}\left(  \gamma^{\mu}D_{\mu}^{f}\right)
\psi,\text{ }\mathcal{O}_{D+2,4}^{f}=\frac{m^{2}}{\hbar}\bar{\psi}\psi.
\end{gather}
For the scalar field $\phi$, the operator with $n=D+2$ which are
gauge-invariant and covariant is
\begin{equation}
\mathcal{O}_{D+2}^{s}=m\phi^{+}\left(  D^{\mu}D_{\mu}\right)  \phi.
\end{equation}
If we truncate the scalar effective theory at $\mathcal{O}\left(  \frac
{1}{\Lambda}\right)  $, by redefining the scalar field $\phi$, it is easy to
see that the truncated effective theory is equivalent to the original field
theory with redefined mass. So we need effective operators at $\mathcal{O}%
\left(  \frac{1}{\Lambda^{2}}\right)  $ to produce nontrivial results. The
basis of independent effective operators with $n=D+3$ is%
\begin{gather}
\mathcal{O}_{D+3,1}^{s}=\hbar^{2}\phi^{+}\left(  D^{\mu}D_{\mu}D^{\nu}D_{\nu
}\right)  \phi,\text{ }\mathcal{O}_{D+3,2}^{s}=\hbar^{2}\phi^{+}\left(
D^{\mu}D^{\nu}D_{\nu}D_{\mu}\right)  \phi,\nonumber\\
\mathcal{O}_{D+3,3}^{s}=\hbar^{2}\phi^{+}\left(  D^{\mu}D^{\nu}D_{\mu}D_{\nu
}\right)  \phi,\text{ }\mathcal{O}_{D+3,4}^{s}=m^{2}\phi^{+}\left(  D^{\mu
}D_{\mu}\right)  \phi,\text{ }\mathcal{O}_{D+3,5}^{s}=\frac{m^{4}}{\hbar^{2}%
}\phi^{+}\phi.
\end{gather}
Define the action $S_{eff}^{s\left(  f\right)  }=\int d^{D+1}x\sqrt{\left\vert
g\right\vert }\mathcal{L}_{eff}^{s\left(  f\right)  }$. Varying $S_{eff}%
^{s\left(  f\right)  }$ with respect to $\phi^{+}\left(  \bar{\psi}\right)  $
gives the deformed KG(Dirac) equation of $\phi\left(  \psi\right)  $. Thus,
the deformed KG equation to $\mathcal{O}\left(  \frac{1}{\Lambda^{2}}\right)
$ is
\begin{gather}
-D^{\mu}D_{\mu}\phi-\frac{m^{2}}{\hbar^{2}}\phi+C_{D+2}^{s}\frac{m}{\Lambda
}D^{\mu}D_{\mu}\phi\nonumber\\
+\frac{C_{D+3,1}^{s}\hbar^{2}}{\Lambda^{2}}D^{\mu}D_{\mu}D^{\nu}D_{\nu}%
\phi+\frac{C_{D+3,2}^{s}\hbar^{2}}{\Lambda^{2}}D^{\mu}D^{\nu}D_{\nu}D_{\mu
}\phi\nonumber\\
+\frac{C_{D+3,3}^{s}\hbar^{2}}{\Lambda^{2}}D^{\mu}D^{\nu}D_{\mu}D_{\nu}%
\phi+\frac{C_{D+3,4}^{s}m^{2}}{\Lambda^{2}}D^{\mu}D_{\mu}\phi+\frac
{C_{D+3,5}^{s}}{\Lambda^{2}}\frac{m^{4}}{\hbar^{2}}\phi=0,
\label{eq:deformedKG}%
\end{gather}
and the deformed Dirac equation to $\mathcal{O}\left(  \frac{1}{\Lambda
}\right)  $ is%
\begin{gather}
i\gamma^{\mu}D_{\mu}^{f}\psi-\frac{m}{\hbar}\psi+\frac{C_{D+2,1}^{f}\hbar
}{\Lambda}\left(  \gamma^{\nu}D_{\nu}^{f}\right)  \left(  \gamma^{\mu}D_{\mu
}^{f}\right)  \psi\nonumber\\
+\frac{C_{D+2,2}^{f}\hbar}{\Lambda}D^{f,\mu}D_{\mu}^{f}\psi+\frac
{iC_{D+2,3}^{f}m}{\Lambda}\left(  \gamma^{\mu}D_{\mu}^{f}\right)  \psi
+\frac{C_{D+2,4}^{f}}{\Lambda}\frac{m^{2}}{\hbar}\psi=0.
\label{eq:deformedDirac}%
\end{gather}

\subsection{Deformed Scalar Hamilton-Jacobi Equation}

To find the classical action $I$ by using WKB approximation to solve eqn.
$\left(  \ref{eq:deformedKG}\right)  $, we make the ansatz for $\phi$ as
before
\begin{equation}
\phi=\exp\left(  \frac{iI}{\hbar}\right)  .
\end{equation}
Substituting it into eqn. $\left(  \ref{eq:deformedKG}\right)  $, one expands
eqn. $\left(  \ref{eq:deformedKG}\right)  $ in powers of $\hbar$ and finds to
the lowest order%
\begin{equation}
A\left(  \partial^{\mu}I+eA^{\mu}\right)  \left(  \partial_{\mu}I+eA_{\mu
}\right)  -Bm^{2}+\frac{C}{\Lambda^{2}}\left[  \left(  \partial^{\mu}%
I+eA^{\mu}\right)  \left(  \partial_{\mu}I+eA_{\mu}\right)  \right]  ^{2}=0,
\label{eq:deformedHamilton-JacobiS}%
\end{equation}
where $A=1-\frac{C_{D+2}^{s}m}{\Lambda}-\frac{C_{D+3,4}^{s}m^{2}}{\Lambda^{2}%
}$, $B=1-\frac{C_{D+3,5}^{s}m^{2}}{\Lambda^{2}}$, and $C\equiv C_{D+3,1}%
^{s}+C_{D+3,2}^{s}+C_{D+3,3}^{s}$. Solving eqn. $\left(
\ref{eq:deformedHamilton-JacobiS}\right)  $ gives%
\begin{equation}
\left(  \partial^{\mu}I+eA^{\mu}\right)  \left(  \partial_{\mu}I+eA_{\mu
}\right)  =m_{eff,\pm}^{2}, \label{eq:deformedHamilton-JacobiSs}%
\end{equation}
where%
\begin{equation}
m_{eff,\pm}^{2}\equiv\frac{-A\pm\sqrt{A^{2}+4BCm^{2}\Lambda^{-2}}}{2C}%
\Lambda^{2}.
\end{equation}
When $m/\Lambda\rightarrow0$, we find $m_{eff,+}^{2}\sim m^{2}$ and
$m_{eff,-}^{2}\sim-\frac{2}{C}\Lambda^{2}$. As $\Lambda\gg m$, eqn. $\left(
\ref{eq:deformedHamilton-JacobiSs}\right)  $ gives $\partial^{\mu}I\sim
\Lambda$ for $m_{eff,-}^{2}$ and hence the action $I$ highly oscillates in
spacetime. One may argue such action is not physical and hence could be
discarded by using the low-momentum consistency condition proposed in
\cite{Ching2012}. Alternatively, one wants to recover the original results
when $\Lambda\rightarrow\infty$. However, as $\Lambda\rightarrow\infty$, the
solution to in eqn. $\left(  \ref{eq:deformedHamilton-JacobiSs}\right)  $ with
$m_{eff,-}^{2}$ blows up while eqn. $\left(
\ref{eq:deformedHamilton-JacobiSs}\right)  $ with $m_{eff,+}^{2}$ becomes eqn.
$\left(  \ref{eq:Hamilton-JacobiS}\right)  $ since $m_{eff,+}^{2}\rightarrow
m^{2}$. Therefore, we pick $m_{eff,+}^{2}$ in eqn. $\left(
\ref{eq:deformedHamilton-JacobiSs}\right)  $. Comparing eqn. $\left(
\ref{eq:deformedHamilton-JacobiSs}\right)  $ with eqn. $\left(
\ref{eq:Hamilton-JacobiS}\right)  $, we find that all the quantum gravity
contributions to the deformed scalar Hamilton-Jacobi equation are included in
one effective parameter, $m_{eff}^{2}$.

\subsection{Deformed Fermionic Hamilton-Jacobi Equation}

To obtain the Hamilton-Jacobi equation for the classical action, the ansatz
for $\psi$ takes the form of
\begin{equation}
\psi=\exp\left(  \frac{iI}{\hbar}\right)  v, \label{eq:fermionansatzD}%
\end{equation}
where $v$ is a vector function of the spacetime. Substituting eqn. $\left(
\ref{eq:fermionansatzD}\right)  $ into eqn. $\left(  \ref{eq:deformedDirac}%
\right)  $, we find to the lowest order of $\hbar$%
\begin{equation}
-A\gamma_{\mu}\left(  \partial^{\mu}I+eA^{\mu}\right)  v=\left[  Bm+\frac
{C}{\Lambda}\left(  \partial^{\mu}I+eA^{\mu}\right)  \left(  \partial_{\mu
}I+eA_{\mu}\right)  \right]  v, \label{eq:deformedHamilton-JacobiF}%
\end{equation}
where $A=1+\frac{C_{D+2,3}^{f}m}{\Lambda}$, $B=1-\frac{C_{D+2,4}^{f}m}%
{\Lambda}$, and $C\equiv C_{D+2,1}^{f}+C_{D+2,2}^{f}.$Multiplying both sides
of eqn. $\left(  \ref{eq:deformedHamilton-JacobiF}\right)  $ from the left by
$-A\gamma_{\nu}\left(  \partial^{\nu}I+eA^{v}\right)  \,$and using eqn.
$\left(  \ref{eq:deformedHamilton-JacobiF}\right)  $ on RHS, we obtain%
\begin{equation}
A^{2}\left(  \partial^{\mu}I+eA^{\mu}\right)  \left(  \partial_{\mu}I+eA_{\mu
}\right)  =\left[  Bm+\frac{C}{\Lambda}\left(  \partial^{\mu}I+eA^{\mu
}\right)  \left(  \partial_{\mu}I+eA_{\mu}\right)  \right]  ^{2}.
\label{eq:deformedHamilton-JacobiFs}%
\end{equation}
Solving eqn. $\left(  \ref{eq:deformedHamilton-JacobiFs}\right)  $, we also
find%
\begin{equation}
\left(  \partial^{\mu}I+eA^{\mu}\right)  \left(  \partial_{\mu}I+eA_{\mu
}\right)  =m_{eff}^{2}, \label{eq:deformedHamilton-Jacobi}%
\end{equation}
where $m_{eff}^{2}$ is a function of $m$, $\Lambda$, $A$, $B$, and $C$ and is
chosen as $m_{eff}^{2}\rightarrow m^{2}$ as $\Lambda\rightarrow\infty$.
Similarly, all the quantum gravity contributions to the deformed fermionic
Hamilton-Jacobi equation are included in one effective parameter, $m_{eff}%
^{2}$.

\subsection{Deformed Hamilton-Jacobi Equation to All Orders}

We have shown that the deformed scalar(fermionic) Hamilton-Jacobi Equation can
be written as the form of eqn. $\left(  \ref{eq:deformedHamilton-Jacobi}%
\right)  $ up to the order of $\frac{1}{\Lambda^{2}}\left(  \frac{1}{\Lambda
}\right)  $. Here, we will show that the deformed scalar(fermionic)
Hamilton-Jacobi Equation can also be written as the form of eqn. $\left(
\ref{eq:deformedHamilton-Jacobi}\right)  $ to all orders of the effective theories.

For a scalar field, the effective operator $\mathcal{O}_{n,j}^{s}$ must
contain even number of $D_{\mu}$ to be covariant. Since $\mathcal{O}_{n,j}%
^{s}$ contains two $\phi$, we find
\begin{equation}
\mathcal{O}_{n,j}^{s}=\hbar^{2q-2}m^{p}\phi^{+}\mathcal{C}\left(  D_{\mu_{1}%
}\cdots D_{\mu_{2q}}\right)  \phi,
\end{equation}
where integers $p,q\geqslant0$, $2q+p=n-D+1$, $j=\left\{  \mathcal{C}%
,p\right\}  $ and $\mathcal{C}$ denotes any possible way of contracting
$\mu_{1}\cdots\mu_{2q}$ in pair to make $\mathcal{O}_{n,j}^{s}$ covariant.
Therefore, the deformed KG equation to all orders of the effective theory is%
\begin{equation}
-D^{\mu}D_{\mu}\phi-\frac{m^{2}}{\hbar^{2}}\phi+\sum_{j,n>D+1}C_{n,j}^{s}%
\frac{\hbar^{2q-2}m^{p}}{\Lambda^{n-\left(  D+1\right)  }}\mathcal{C}\left(
D_{\mu_{1}}\cdots D_{\mu_{2q}}\right)  \phi=0. \label{eq:deformedKGH}%
\end{equation}
Making the ansatz $\phi=\exp\left(  \frac{iI}{\hbar}\right)  $ and
substituting it into eqn. $\left(  \ref{eq:deformedHamilton-Jacobi}\right)  $
gives the deformed the Hamilton-Jacobi equation for $I$
\begin{equation}
G\left(  X\right)  \equiv X-m^{2}+f\left(  X\right)  =0,
\label{eq:deformedHamilton-JacobiSH}%
\end{equation}
where $X\equiv\left(  \partial^{\mu}I+eA^{\mu}\right)  \left(  \partial_{\mu
}I+eA_{\mu}\right)  $ and
\[
f\left(  x\right)  \equiv\sum_{j,n>D+1}\frac{\left(  -1\right)  ^{q}%
C_{n,j}^{s}m^{p}x^{q}}{\Lambda^{n-\left(  D+1\right)  }}.
\]
When $m/\Lambda\ll1$ as assumed, for eqn. $\left(
\ref{eq:deformedHamilton-JacobiSH}\right)  $, there exists one and only one
root $m_{eff}^{2}\,$\ in $\left(  0,2m^{2}\right)  $. In fact, since
$m/\Lambda\ll1$, $f\left(  0\right)  $ and $f\left(  2m^{2}\right)  \sim
m^{2}\frac{m}{\Lambda}\ll m^{2}$. Thus, one finds $G\left(  0\right)  <0$ and
$G\left(  2m^{2}\right)  >0$. So there exists at one root of eqn. $\left(
\ref{eq:deformedHamilton-JacobiSH}\right)  $ in $\left(  0,2m^{2}\right)  $.
On the other hand, using $f^{\prime}\left(  x\right)  \sim\frac{m}{\Lambda}%
\ll1$ for $x\in\left(  0,2m^{2}\right)  $, one finds is $G^{\prime}\left(
x\right)  >0$ for $x\in\left(  0,2m^{2}\right)  $. This completes the proof of
existence of one and only one root of eqn. $\left(
\ref{eq:deformedHamilton-JacobiSH}\right)  \,$\ in $\left(  0,2m^{2}\right)
$. There might be other roots which is not in $\left(  0,2m^{2}\right)  $.
However, they are not physical and discarded since they don't approach\ $m^{2}%
$\ as $\Lambda\rightarrow\infty$. By solving eqn. $\left(
\ref{eq:deformedHamilton-JacobiSH}\right)  $, we find the classical action $I$
satisfies $\ $%
\begin{equation}
\left(  \partial^{\mu}I+eA^{\mu}\right)  \left(  \partial_{\mu}I+eA_{\mu
}\right)  =m_{eff}^{2},
\end{equation}
where $m_{eff}^{2}$ is uniquely determined by $\Lambda$, $m$, and $C_{n,j}%
^{s}$.

For a fermion field, the effective operator $\mathcal{O}_{n,j}^{f}$ can be
written as%
\begin{equation}
\mathcal{O}_{n,j}^{f}=i^{q}\hbar^{q-1}m^{p}\bar{\psi}\mathcal{C}\left(
D_{\mu_{1}}^{f}\cdots D_{\mu_{q}}^{f}\gamma_{\nu_{1}}\cdots\gamma_{\nu_{2r-q}%
}\right)  \psi,
\end{equation}
where integers $q$, $p$, $r\geqslant0$, $p+q=n-D$, $q\geqslant r\geqslant
\frac{q}{2}$, $j=\left\{  \mathcal{C},q,r\right\}  $ and and $\mathcal{C}$
denotes any possible way of contracting $\mu_{1}\cdots\mu_{q}$ and $\nu
_{1}\cdots\nu_{2r-q}$ in pair. Therefore, the deformed Dirac equation to all
orders of the effective theory is%
\begin{equation}
i\gamma^{\mu}D_{\mu}^{f}\psi-\frac{m}{\hbar}\psi+\sum_{j,n>D+1}C_{n,j}%
^{f}\frac{i^{q}\hbar^{q-1}m^{p}}{\Lambda^{n-\left(  D+1\right)  }}%
\mathcal{C}\left(  D_{\mu_{1}}^{f}\cdots D_{\mu_{q}}^{f}\gamma_{\nu_{1}}%
\cdots\gamma_{\nu_{2r-q}}\right)  \psi=0. \label{eq:deformedDiracH}%
\end{equation}
Substituting the ansatz $\psi=\exp\left(  \frac{iI}{\hbar}\right)  v$ into
eqn. $\left(  \ref{eq:deformedDiracH}\right)  $, we find to the lowest order
of $\hbar$%
\begin{equation}
\left[  1-g\left(  X\right)  \right]  \gamma_{\mu}\left(  \partial^{\mu
}I+eA^{\mu}\right)  v=\left[  f\left(  X\right)  -m\right]  v
\label{eq:deformedHamilton-JacobiFH}%
\end{equation}
where $X\equiv\left(  \partial^{\mu}I+eA^{\mu}\right)  \left(  \partial_{\mu
}I+eA_{\mu}\right)  $ and%
\[
f\left(  x\right)  \equiv\sum_{j,n>D+1,q\text{ is even}}\frac{C_{n,j}^{f}%
m^{p}x^{\frac{q}{2}}}{\Lambda^{n-\left(  D+1\right)  }},\text{ and }g\left(
x\right)  \equiv\sum_{j,n>D+1,q\text{ is odd}}\frac{-C_{n,j}^{f}m^{p}%
x^{\frac{q-1}{2}}}{\Lambda^{n-\left(  D+1\right)  }}.
\]
Using the same manipulation as before, we find that eqn. $\left(
\ref{eq:deformedHamilton-JacobiFH}\right)  $ reduces to%
\begin{equation}
H\left(  X\right)  \equiv\left[  1-g\left(  X\right)  \right]  ^{2}X-\left[
f\left(  X\right)  -m\right]  ^{2}=0. \label{eq:deformedHamilton-JacobiFHs}%
\end{equation}
For $m/\Lambda\ll1$, one can show that $f\left(  0\right)  $, $f\left(
2m^{2}\right)  \ll m$ and $g\left(  0\right)  $, $g\left(  2m^{2}\right)
\ll1$. Thus, $H\left(  0\right)  <0$ and $H\left(  2m^{2}\right)  >0$.
Moreover, $H^{\prime}\left(  x\right)  >0$ for $x\in\left(  0,2m^{2}\right)  $
since $f^{\prime}\left(  x\right)  \ll1$ and $xg^{\prime}\left(  x\right)
\ll1$ for $x\in\left(  0,2m^{2}\right)  $. Therefore, there exists one and
only one root $m_{eff}^{2}\,$\ of eqn. $\left(
\ref{eq:deformedHamilton-JacobiFHs}\right)  $ in $\left(  0,2m^{2}\right)  $.
Eqn. $\left(  \ref{eq:deformedHamilton-JacobiFHs}\right)  $ leads to $\ $%
\begin{equation}
\left(  \partial^{\mu}I+eA^{\mu}\right)  \left(  \partial_{\mu}I+eA_{\mu
}\right)  =m_{eff}^{2},
\end{equation}
where $m_{eff}^{2}$ is uniquely determined by $\Lambda$, $m$, and $C_{n,j}%
^{f}$.

\section{Discussion and Conclusion}

\label{Sec:Con}

The Hamilton-Jacobi method without and with the incorporation of the quantum
gravity effects has been studied in this paper. Specifically, we have
calculated the scalar and fermionic Hamilton-Jacobi equations for the
classical action $I$ in the background of a $\left(  D+1\right)  $-dimensional
black hole with the metric $g_{\mu\nu}$ and the electromagnetic potential
$A_{\mu}$. First, in Section \ref{Sec:HJ}, we revisited the derivation of the
original Hamilton-Jacobi equations for the action $I$ of tunneling of scalar
and fermionic particles from the black hole. In the framework of effective
field theories constructed in Section \ref{Sec:HJG}, the deformed
model-independent KG/Dirac equations respecting covariance of spacetime and
gauge invariance of $A_{\mu}$ have then been derived. Finally, in Section
\ref{Sec:HJG}, substituting the WKB ansatz for the scalar and fermionic
wavefunctions into the deformed KG/Dirac equations, we expanded them in powers
of $\hbar$, kept only the lowest order and hence gave the deformed
Hamilton-Jacobi equations.

Our results are summarized as follows:

\begin{enumerate}
\item[$\left(  a\right)  $] In the case of no quantum gravity effects, we have
shown in Section \ref{Sec:HJ} that the fermionic Hamilton-Jacobi equation for
the action $I$ can be written in the same form of the scalar one. Both can be
written as
\begin{equation}
\left(  \partial^{\mu}I+eA^{\mu}\right)  \left(  \partial_{\mu}I+eA_{\mu
}\right)  =m^{2}, \label{eq:HJeqn}%
\end{equation}
where $A^{\mu}\ $is the black hole's electromagnetic potential and $m$ is the
mass of the particle.

\item[$\left(  b\right)  $] In the case of covariant low energy quantum
gravity effective models, we have shown in Section \ref{Sec:HJG} both scalar
and fermionic deformed Hamilton-Jacobi equations can be reduced to
\begin{equation}
\left(  \partial^{\mu}I+eA^{\mu}\right)  \left(  \partial_{\mu}I+eA_{\mu
}\right)  =m_{eff}^{2}, \label{eq:HJeqnD}%
\end{equation}
where all the quantum gravity contributions are included in only one parameter
$m_{eff}^{2}$.
\end{enumerate}

As a bonus of Result $\left(  a\right)  $, it provides a shortcut to calculate
the action $I$ of tunneling of a fermionic particle from the black hole.
Instead of solving the complicated matrix equation $\left(
\ref{eq:Hamilton-JacobiF}\right)  $, we can solve eqn. $\left(  \ref{eq:HJeqn}%
\right)  $ for $I$. For example, such shortcut was discussed in the case of
fermion tunneling from the Bardeen-Vaidya black hole\cite{Criscienzo2008}.
Since both scalar and fermion actions satisfy the same equation $\left(
\ref{eq:HJeqn}\right)  $, the Hamilton-Jacobi method relating the imaginary
part of the actions to Hawking temperature indicates that Hawking temperature
for scalar and fermion particles are the same. In fact, using the
Hamilton-Jacobi method, Hawking temperatures were calculated for a scalar and
a fermion, in the context of charged BTZ black holes\cite{Ejaz2013}, black
strings\cite{Ahmed2011,Gohar2013}, \textit{etc}., and found the same results.
As shown above, the coincidence is guaranteed by eqn. $\left(  \ref{eq:HJeqn}%
\right)  $.

When incorporating the quantum gravity into quantum field theory, as mentioned
in Introduction, there are two kinds of effective models, one of which
respects covariance and the other does not. For the non-covariant effective
models, the deformed Hamilton-Jacobi method with inclusion of quantum gravity
effects were studied in
\cite{Chen:2013pra,Chen2014-1,Chen2014-2,Chen:2013kha,Ren2014,Chen:2014xsa}.
There, it has been shown that the corrections to the original Hawking
temperature depend on the quantum numbers of the emitted particles in a
non-trivial way. In some cases, such corrections could lead to the remnant
left in the evaporation\cite{Chen:2013pra,Chen2014-1,Chen2014-2}. In this
paper, it is first time to investigate the deformed Hamilton-Jacobi method in
covariant effective models. In this scenario, the only difference between the
deformed Hamilton-Jacobi equations $\left(  \ref{eq:HJeqnD}\right)  $ and the
original ones $\left(  \ref{eq:HJeqn}\right)  $ is the parameters on RHS. As a
consequence, if the original Hawking temperature of some black hole is
independent of the mass of emitted particles, there will no corrections from
the quantum gravity. Even if the original Hawking temperature depends on the
mass $m$, the corrected Hawking temperature can be obtained from the original
one by simply replacing $m$ by $m_{eff}$. In either case above, if a black
hole evaporates by the original Hawking radiations, it will also evaporate by
the deformed Hawking radiations. Therefore, one may argue that covariance of
spacetime in quantum gravity effective models has to be broken in order for a
black hole to have the remnant left in the evaporation.

\begin{acknowledgments}
We are grateful to Houwen Wu and Zheng Sun for useful discussions. This work
is supported in part by NSFC (Grant No. 11005016, 11175039 and 11375121) and
SYSTF (Grant No. 2012JQ0039).
\end{acknowledgments}

\end{document}